\documentclass[twocolumn,aps,prl,showpacs,amsmath,superscriptaddress]{revtex4}
\usepackage{amssymb}
\usepackage{mathrsfs}
\usepackage{graphicx}
\usepackage{float}

\begin{document}

\title{Fractional Topological States of Dipolar Fermions in One-Dimensional Optical Superlattices}

\author{Zhihao Xu}
\author{Linhu Li}
\author{Shu Chen}
\email{schen@aphy.iphy.ac.cn}\affiliation{Beijing National
Laboratory for Condensed Matter Physics, Institute of Physics,
Chinese Academy of Sciences, Beijing 100190, China}
\date{ \today}

\begin{abstract}
We study the properties of dipolar fermions trapped in
one-dimensional bichromatic optical lattices and show the
existence of fractional topological states in the presence of
strong dipole-dipole interactions. We find some interesting
connections between fractional topological states in
one-dimensional superlattices and the fractional quantum Hall
states: (i) the one-dimensional fractional topological states for
systems at filling factor $\nu=1/p$ have $p$-fold degeneracy,
(ii) the quasihole excitations fulfill the same counting
rule as that of fractional quantum Hall states, and (iii)
the total Chern number of $p$-fold degenerate states is a nonzero
integer.
The existence of crystalline order in our system is also consistent with the thin-torus limit
of the fractional quantum Hall state on a torus.
%distinguishes it from the fractional quantum Hall state.
The possible experimental realization in cold atomic systems offers a
new platform for the study of fractional topological phases in
one-dimensional superlattice systems.
\end{abstract}
%05.30.Fk Fermion systems and electron gas
%03.75.Hh Static properties of condensates; thermodynamical,
%statistical, and structural properties
%73.21.Cd Superlattices
%71.10.Fd,%Lattice fermion models (Hubbard model, etc.)

\maketitle

%\section{Introduction}
{\it Introduction.-} Fractional quantum Hall (FQH) effects have
attracted intensive studies in the past decades as an important
subject in condensed matter physics. The traditional FQH states
were realized in two-dimensional (2D) electron gases within a
strong external magnetic field. In addition to 2D electron gases,
great effort has been made to study quantum Hall effects in some
other physical systems, for example, lattice systems without a
magnetic field and cold atomic systems. Effective Landau levels
can be realized in cold atomic systems in the presence of a
rapidly rotating trap \cite{Cooper, Fetter} or a laser-induced
gauge field \cite{Spielman}. Because of the existence of
long-range interaction, the dipolar Fermi gas is a good candidate
to realize FQH states. The FQH effects in a 2D dipolar Fermi
gas with either isotropic \cite{Osterloh07} or anisotropic
dipole-dipole interaction (DDI) \cite{Qiu} have been studied recently.

As most of the previous studies on topological nontrivial states focus
on 2D systems \cite{Osterloh05,Shao,Kane}, the one-dimensional
(1D) systems attracted less attention until very recently
\cite{Lang,Kraus,Zhu}. Although 1D systems without any symmetry
are generally topologically trivial, the 1D
superlattice model was recently found to be topologically nontrivial
\cite{Lang,Kraus} as the periodic parameter in these
superlattice models can be considered as an additional dimension
and thus the systems may have a nontrivial Chern number in an
effective 2D parameter space.
%The single-particle spectrum of a 1D
%quasi-periodic lattice is organized in bands.
It has been shown that the 1D superlattice system with subbands
being filled is not a trivial band insulator but a topological
insulator with a nonzero Chern number \cite{Lang}, which can be
viewed as a correspondence of the integer quantum Hall state of a
2D square lattice \cite{TKNN,Hofstadter} in the reduced 1D system.
It is well known that the FQH effect emerges
from the integer quantum Hall state in the presence of strong
long-range interactions. It is natural to ask whether a fractional
topological state is available for the 1D superlattice system
when interactions are included.

In this Letter, we explore the nontrivial topological properties of
dipolar fermions in 1D bichromatic optical lattices,
which can be realized in cold atom experiments by loading the
dipolar fermions into the lattice superposed by two 1D optical
lattices with different wavelengths \cite{exp1,exp2}. The
noninteracting part of our Hamiltonian is the recently studied 1D
superlattice model with topologically nontrivial bands
\cite{Lang,Kraus}. The presence of dipolar interactions
breaks down the band description within a noninteracting picture.
To characterize topological features of the interacting system, we
study the low-energy spectrum and the topological Chern number of
the dipolar system based on exact calculation of finite-size
systems. The existence of nontrivial topological states for the
strongly interacting system at fractional filling is demonstrated
by the topological degeneracy and nontrivial Chern number of the
low-energy states. Particularly, recent progress in manipulating
ultracold polar molecules \cite{dipolar} offers the possibility of
exploring exotic quantum states of Fermi gases with strong dipolar
interactions in the topologically nontrivial optical
lattices.

%\section{Model Hamiltonian}
{\it Model Hamiltonian.-} We consider a 1D Fermi gas with
DDIs in a bichromatic optical
superlattice:
\begin{equation}
\label{eqn1}
H=-t\sum_{i}{(c_{i}^{\dag}c_{i+1}+\mathrm{H.c.})}+\sum_{i}{\mu_{i}n_i}+\frac{V}{2}\sum_{i\neq
j}{\frac{n_i n_j}{|i-j|^3}}
\end{equation}
with
\begin{equation}
\label{eqn2} \mu_{i}=\lambda \cos{(2\pi\alpha i+\delta)},
\end{equation}
where $c_{i}^{\dag}$ ($c_{i}$) is the creation (annihilation)
operator of fermions, $n_i=c_{i}^{\dag}c_{i}$ the density
operator, and $t$ the hopping strength.
Here $\mu_i$ is the periodic potential with
$\lambda$ being the modulation amplitude,
$\alpha$ determining the modulation period and $\delta$ being an
arbitrary phase. The last term of Eq. (\ref{eqn1}) is for DDIs
which are long-range interactions decayed with $1/r^3$ with $V$
the strength of DDI. For convenience, we shall set $t=1$ as the unit
of energy and choose $\alpha=1/q$.
%so the 1D superlattice has a unit cell of $q$ sites.
%and the
%single-particle spectrum is composed of $q$ bands.

In the absence of interactions, it has been demonstrated that the
system with its subbands fully filled by fermions is an
insulator with a nontrivial topological Chern number in a 2D
parameter space spanned by momentum and the phase of $\delta$
\cite{Lang}.
%for example, the Chern number for the state with the
%lowest sub-band filled is $1$.
If the subband is only partially
filled, the system is a topologically trivial conductor. In this
Letter, we shall study the case with the lowest band being partially
filled by fermions subjected to the long-range interaction. Given
that the number of fermions is $N$ and the lattice size
is $L$, the filling factor is defined as $\nu=N/N_{\mathrm{cell}}$ with
$N_{\mathrm{cell}}=L / q$ being the number of primitive cells. While
$\nu=1$ corresponds to the lowest band being fully filled,
in this Letter we shall consider the system with a
fractional filling factor, for example, $\nu=1/3$ and $\nu=1/5$.

\begin{figure}[tbp]
\includegraphics[height=9cm,width=10cm] {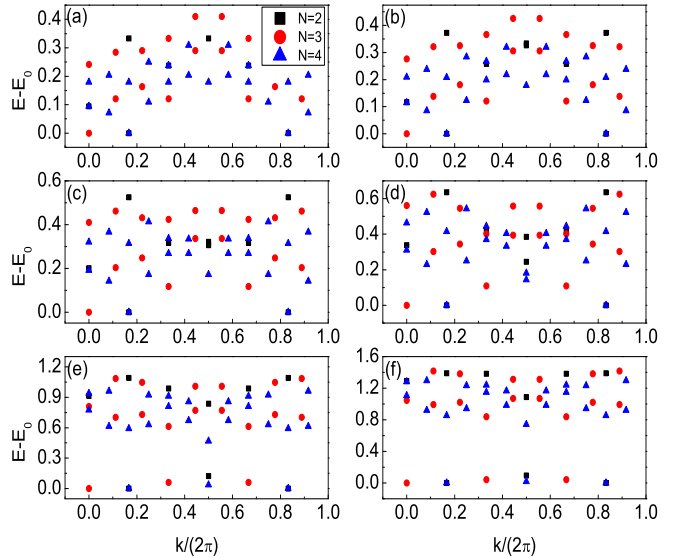}
\caption{(Color online) Low-energy spectrum in momentum space with
filling factor $\nu=1/3$, $t=1$, $\lambda=1.5$, $\alpha=1/3$, $\delta=5\pi/4$
and different $V$ under the PBC. (a)-(f) $V$ takes 0, 1,
10, 50, 300, and 500, respectively.}\label{Fig1}
\end{figure}

%\section{Results}
%\subsection{Low-energy spectrum and ground state degeneracy}
{\it Low-energy spectrum and ground state degeneracy.-} In the
presence of the long-range DDI term, we diagonalize the
Hamiltonian (\ref{eqn1}) in each momentum subspace with $k=2\pi
m/N_{\mathrm{cell}}$ under the periodic boundary condition (PBC) \cite{suppl}, where
$m$ takes $0,1,\ldots,N_{\mathrm{cell}}-1$. For
finite-size systems with a given fractional filling, we study the
change of low-energy spectrum with the increase in the interaction strength
$V$. In Fig.\ref{Fig1}, we display the
low-energy spectrum in momentum sectors for systems with
$t=1$, $\nu=1/3$, $\lambda=1.5$, $\alpha=1/3$, $\delta=5\pi/4$, and
different $V$. Various cases with particle numbers $N=2,3,4$ are
shown in the same figure.
When $V$ is small, it is hard to distinguish the lowest states apart from the
higher excited states by an obvious gap.  When $V$ exceeds $50$,
the lowest three states tend to form a ground-state (GS) manifold with
an obvious gap separating them from higher ones. As $V$
increases further, the gap becomes more obvious and the lowest
three states become nearly degenerate at $V=500$. The threefold
degeneracy does not depend on particle numbers, but is only
relevant to the filling factor $\nu=1/3$.

In the large $V$ case, the lowest three states in the GS
manifold always appear at some determinate positions in the
momentum space. For cases of $N=2,4$ (even), the total momenta
locate at $K=\pi/3,~\pi,~5\pi/3$, whereas for $N=3$ (odd), at
$K=0,~2\pi/3,~4\pi/3$. We observe that one
can make a connection between the momenta in our system and the
orbital momenta of the FQH system in the thin-torus limit by setting
$N_{\phi}=N_{\mathrm{cell}}$ with $N_{\phi}$ being the number of the flux
quanta.
According to the exclusion rule known from the thin-torus limit of the
FQH system \cite{Regnault,Bergholtz},
the total momenta of the $p$-fold degenerate GSs emerge at
$K=(2\pi)\{[p N(N-1)/2+l N]\mathrm{mod}N_{\mathrm{cell}}\}/N_{\mathrm{cell}}$, where
$l = 0,1,\cdots,p-1$ for a system with $\nu=1/p$.
%Similar to the FQH, applying a generalized Pauli principle
%\cite{Regnault}, that there are no more than one particle in the
%$3$ successive orbitals for $\nu=1/3$, consequently, there are $3$
%cases for the occupation numbers $100100\ldots 100$, $010010\ldots
%010$ and $001001\ldots 001$ and the lowest $3$-fold states emerge
%at $K=(2\pi)\{[3N(N-1)/2+l_1N]\mathrm{mod}N_{cell}\}/N_{cell}$,
%where $l_1=0,1,2$.
We also check the low-energy spectra for
systems with $t=1$, $\nu=1/5$, $\lambda=1.5$, $\alpha=1/3$,
$\delta=5\pi/4$, and different $V$. Similar to cases of $\nu=1/3$,
both systems with $N=2$ and $N=3$ show the fivefold degenerate
GS manifold in the large $V$ limit and the positions of momenta
fulfill the above expression determined by the exclusion rule.
%The difference is
%that one needs a larger $V$ to ensure the system to reach the
%regime of nearly degenerate GS manifold as a result of
%the lower filling factor.
%%Similar analysis of the generalized
%%Pauli principle can be applied to the case of $\nu=1/5$, for which
%%the momenta of the 5-fold degenerate ground states emerge at
%%$K=(2\pi)\{[5N(N-1)/2+l_2N]\mathrm{mod}N_{cell}\}/N_{cell}$, where
%%$l_2=0,1,2,3,4$.
%The appearance of $p$-fold degeneracy ($p=1/\nu$
%for $\nu=1/3$ and $\nu=1/5$) reminds us its similarity to the FQH
%system, which has $p$-fold degenerate ground states with a finite
%energy gap separating the ground-state manifold from the excited
%states.

\begin{figure}[tbp]
\includegraphics[height=8cm,width=9cm] {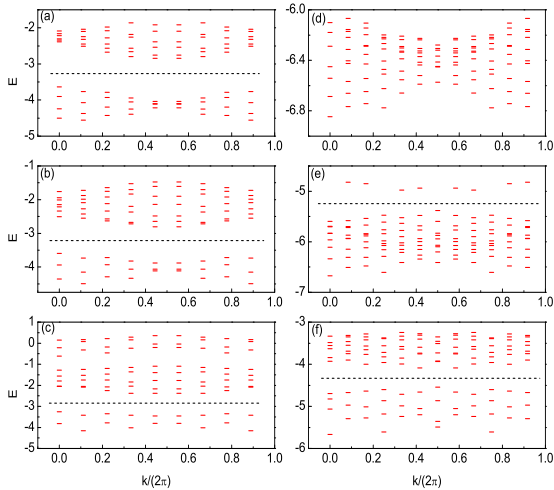}
\caption{(Color online) Quasihole excitation spectrum. The left part
is for the system with $N=2$, $L=27$ and $V=1, ~50, ~500$ from top
to bottom. The right part is for $N=3$, $L=36$ and $V=1, ~50, ~500$
from up to down. }\label{Fig2}
\end{figure}

%\subsection{Quasihole excitation spectrum}
{\it Quasihole excitation spectrum.-} For FQH states, the
existence of quasihole excitations, which fulfill fractional
statistics \cite{Laughlin,Halperin},
is an important characteristic feature of the system. According to
the general counting rule \cite{Regnault}, the number of
quasiholes for the FQH system of $\nu=1/3$ reads as
%\begin{equation*}
%\label{eqn3}
%N_{qh}=N_{cell}\frac{(N_{cell}-2N-1)!}{N!(N_{cell}-3N)!}.
%\end{equation*}
$N_{qh}=N_{\mathrm{cell}}\frac{(N_{\mathrm{cell}}-2N-1)!}{N!(N_{\mathrm{cell}}-3N)!}$.
%which is for the case of $q=3$.
Next we study the quasihole excitations by removing a particle
from our system and check whether a similarity to the FQH system
exists. On the left part of Fig. \ref{Fig2}, we show the
quasihole excitation spectra for the system with $N=2$ and
$L=27$ produced by removing a particle from the system of $\nu=1/3$
with $N=3$ and $L=27$, whereas the right part of Fig.\ref{Fig2}
gives spectra for $N=3$ and $L=36$ by
removing a particle from the system of $\nu=1/3$ with $N=4$ and
$L=36$. For both parts, from top to bottom $V=1$, $50$, and
$500$, respectively.
In the regime of small $V$, the number of quasihole
excitation is much larger than that given according to the above
accounting rule of FQH systems.
As $V$ increases, the low-energy
parts are excited into the upper part.
For $V=500$, as shown in
Fig.\ref{Fig2}(c), below the gap, the number of quasihole excitations
is $18$ with two states on each momentum sector. In Fig.\ref{Fig2}(f),
the total number of states under the dash line
is $40$ and for momentum sectors $[kN_{\mathrm{cell}}/(2\pi)]\mathrm{mod}3=0$,
the number of states below the gap is four while in the
other parts is three, due to the finite-size effect.
For both cases with $V=500$, the total number of states below the
gap is consistent with the number obtained by the counting rule
for the $\nu=1/3$ FQH state.

%\subsection{Topological feature of ground-state manifold}
{\it Topological feature of ground-state manifold.-} To
characterize the topological feature of the many-body states,
we introduce the twist boundary condition
$\psi(r + L,\delta)=e^{i\theta}\psi(r,\delta)$, where $\theta$ is
the introduced phase factor.
%which can be effectively viewed as
%introduction of a magnetic field of $\theta/2\pi$ flux quantum
%threading through the ring.
Under the twist boundary condition,
the momentum $k$ in Brillouin zone gets a shift $k=(2\pi m+\theta)
/N_{\mathrm{cell}}$ with $m=0,1,\ldots ,N_{\mathrm{cell}}-1$. Correspondingly, the
energies vary continuously with the change of $\theta$.  In Fig.3,
we show the low energy spectra as a function of
$\theta$ (the spectrum flux) at a fixed $\delta=5\pi/4$ for
systems with $N=2$ [(a)-(c)] and $N=3$ [(d)-(f)]. In the small $V$
regime of $V=1$, the lower energy
levels overlap together with the change of $\theta$.
For $V=50$, the lowest three energy spectra flow into each other
but are already separated from the higher states. For
$V=500$, the lowest three states are nearly
degenerate, and the GS manifold is well separated from
the higher states by a gap. Similarly, if the phase
$\delta$ varies from $0$ to $2\pi$, the spectrum for a given
$\theta$ changes continuously with the GS manifold well
separated from the other states, which indicates the robustness of
GS manifold in the large $V$ regime. An example for $\theta=0$
corresponding to the PBC is given in Fig.4(a).
\begin{figure}
\includegraphics[height=9cm,width=9cm] {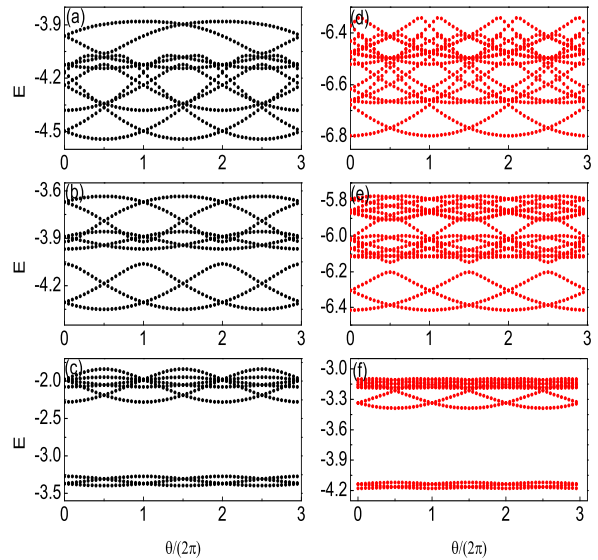}
\caption{(Color online) Spectrum flux versus $\theta$ for systems
with $t=1$, $\nu=1/3$, $\lambda=1.5$, $\alpha=1/3$, $\delta=5\pi/4$ and
different $V$. (a)- (c) $N=2$, $V=1$, $50$, and $500$,
respectively. (e)-(f) $N=3$, $V=1$, $50$, and $500$,
respectively. }\label{Fig3}
\end{figure}

In the 2D parameter space of $(\theta,\delta)$, the
Chern number of the many-body state is defined as an integral
invariant $C=\frac{1}{2\pi}\int{d\theta d\delta
F(\theta,\delta)}$, where $F(\theta,\delta) = \mathrm{Im}(\langle
\frac{\partial \psi}{\partial \delta}| \frac{\partial
\psi}{\partial\theta}\rangle -\langle \frac{\partial
\psi}{\partial \theta}| \frac{\partial\psi}{\partial\delta}
\rangle)$ is the Berry curvature \cite{TKNN,Niu}. Considering the
system with $V=500$ shown in Fig.1(f), we calculate the Chern
numbers of the lowest three nearly degenerate states in the
GS manifold. For system of $N=2$, the Chern numbers of
the three states are $C_1=0.4036$, $C_2=0.1928$, and $C_3=0.4036$,
respectively. For each state, the Chern number is not an integer,
but their sum is an integer
$\sum_{i=1}^3 C_{i}= 1$. Similarly, for $N=3$, we have
$C_1=0.2776$, $C_2=0.4448$, and $C_3=0.2776$ with their summation
being $1$. The existence of a nonzero total Chern
number characterizes the system at
fractional filling having nontrivial topological properties.
Effectively, the total Chern number is shared by the $q$
degenerate states, which is similar to the FQH system with its
$q$-fold GSs sharing an integer total Chern number
\cite{Sheng}.
\begin{figure}
\includegraphics[height=8cm,width=8cm] {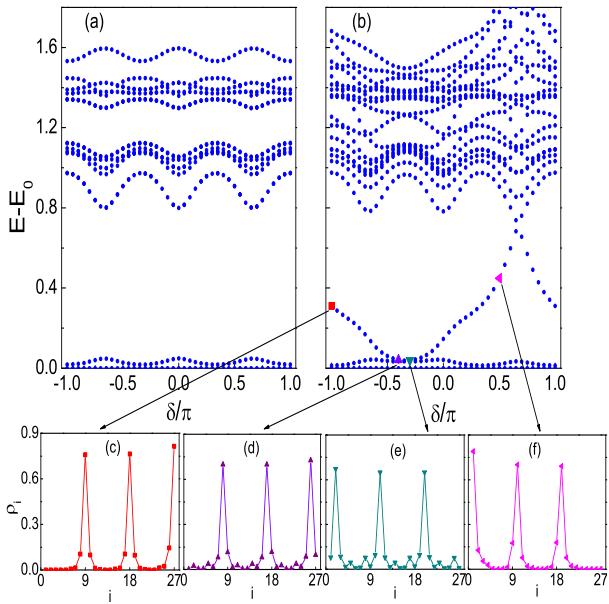}
\caption{(Color online) The low energy spectrum as the function of the phase $\delta$
for the system with $L=27$, $N=3$, $t=1$, $\lambda=1.5$, $\alpha=1/3$, $V=500$.
Here $E_0$ represents the GS energy of the system.
(a) is for the PBC; (b) is for the OBC; (c)-(f) is the density distribution for the
in-gap mode with $\delta=-\pi$, $-0.4\pi$, $-0.3\pi$, and $0.5\pi$, respectively.}\label{Fig4}
\end{figure}

The emergence of edge states under the open boundary condition (OBC) is generally
characteristic of topologically nontrivial phases.
In Fig.4(b), we show the low-energy spectra as a function of phase $\delta$ under the OBC, which is
obtained by setting the hopping amplitude between the first and $L$-th site as zero. In
contrast to the spectra under the PBC, inside the gap regime, there emerge edge modes
which connect the GS and excitation branch of the bulk spectrum as
one varies phase $\delta$.  As shown in the Fig.4(c)-Fig.4(f), the state
is adiabatically changed with $\delta$ varying from $-\pi$ to $\pi/2$.
The density distribution $\rho_i$ shows that the in-gap state with $\delta=-\pi$ is pinned down on
the right edge, whereas the state with $\delta=\pi/2$ is pinned down on the left edge.
\begin{figure}[tbp]
\includegraphics[height=8cm,width=9cm] {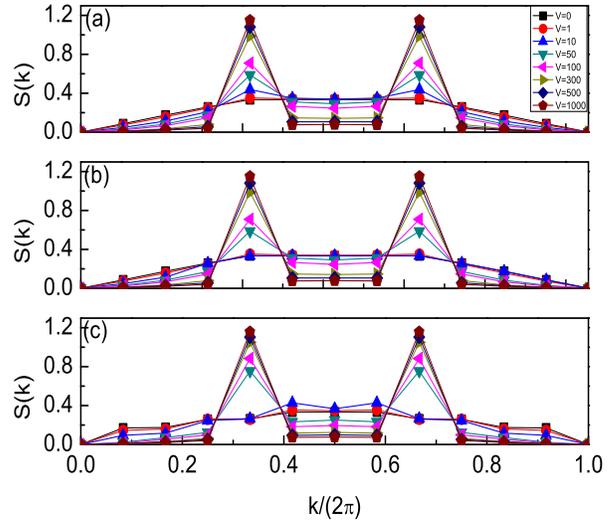}
\caption{(Color online) Static structure factor $S(k)$ of the
lowest three energy states versus momentum $k$ for the system with
different $V$. Here, $N=4,~L=36$.}\label{Fig5}
\end{figure}

%\subsection{Crystallized phase}
%{\it Crystallized phase.-}
The density distributions shown in Fig.4(c)-Fig.4(f) already display the signature
of a crystalized phase.
To reveal the crystalline character of systems with the PBC, we introduce the static structure factor
$S(k)$ defined as
\begin{equation}
\label{eqn4} S(k)=\frac{1}{N_{\mathrm{cell}}}\sum_{i,j}e^{ik(i-j)}[\langle
n_i^c n_j^c \rangle - \langle n_i^c \rangle \langle n_j^c
\rangle],
\end{equation}
where $n_i^c=n_{qi}+n_{qi+1}+\ldots +n_{qi+q-1}$ is the sum of the
particle number operators in the $i$th primitive cell and
$i=0,1,\ldots N_{\mathrm{cell}}-1$. Figure \ref{Fig5} shows the static
structure factor of the three lowest eigenstates versus momentum for
the system with $N=4$, $L=36$, and different $V$.
%In the case of
%small $V$, the $3$-lowest eigenstates announce the different
%behaviors. However, they are assimilation in strong repulsive limits,
In the strong repulsive limit, peaks emerge at $k=2\pi/3,~4\pi/3$
for all the lowest three eigenstates. With the increases of $V$,
the height of the peaks increase and the central parts of the
$S(k)$ decrease dramatically, which suggests that these states are
crystalized with a periodic structure in the large $V$ limit.
%The emergence of crystallized order of dipolar fermions in the
%bichromatic lattice implies that the 1D fractional topological
%state is different from the FQH state, although they have some
%similar topological characters.
The existence of topologically nontrivial crystallized phase in our system is consistent with
previous works on the evolution of FQH states on a torus
\cite{Seidel,Bergholtz-PRL,Bergholtz2}, which have shown that the FQH state
on a torus is adiabatically connected to a crystallized phase as
the 2D system is deformed to the 1D thin-torus limit.

The emergence of fractional topological states is consequence of interplay of nontrivial topology of superlattices and
long-range interactions, which is illustrated by the
shift of edge mode from one to the other edge driven by the phase $\delta$.
While the long-range
interaction is responsible for the formation of degenerate GS manifold, the nontrivial topology of superlattices guarantees
the existence of edge states. Consequently, when the PBC is
changed to the OBC, the threefold degenerate GSs
are lifted and one of them develops into the edge mode as shown in Fig.4(b).
This is quite different from the noninteracting case, for which the GS is nondegenerate
and nontrivial edge states only appear at the integer filling. To see clearly the effect of long-range
interaction, we also check cases with the Coulomb interaction and short-range interactions
\cite{suppl}. While our conclusions also hold true for the case with Coulomb interaction,
no degenerate GS manifold and fractional
topological states are found for the case with short-range interactions even in the strongly interacting limit.

%\section{Summary}
In summary, we demonstrate the existence of
fractional topological states for dipolar fermions in
topologically nontrivial 1D superlattices, which are characterized
by the GS degeneracy, nontrivial total Chern number of
GSs, and quasihole excitations fulfilling the same
counting rule as the FQH states. The existence of crystallized order
in the 1D fractional topological phases is also identified by calculating
the structure factor.
%phase
%However, we also find the
%existence of crystallized order in the 1D fractional topological
%phases,
%which is absent in the 2D FQH states but may appear in the
%thin-torus limit of the quantum Hall system.
Our study provides a
way of creating nontrivial fractional topological states by
trapping the dipolar fermions in 1D bichromatic optical lattices
which are realizable in current cold atomic experiments.

\begin{acknowledgments}
We thank X. Wan and Z. Liu for helpful
discussions. This work has been supported by National Program for
Basic Research of MOST, the NSF of China under Grants
No.11174360 and No.11121063, and 973 grant.
\end{acknowledgments}

\onecolumngrid
\newpage
\section{Supplemental material for ``Fractional topological states of dipolar fermions in
one-dimensional optical superlattices"}

\subsection{I. Exact diagonalization of Hamiltonian in momentum
space} In the supplemental material, we give details for exact
diagonalization of Hamiltonian (1) considered in the main text.
The Hamiltonian is given by
\begin{equation}
\label{eqn1} H=-t\sum_{i}{(c_{i}^{\dag}c_{i+1}+\mathrm{H.c.})}+
\sum_{i}{\lambda \cos{(2\pi\alpha i+\delta)}n_i}+
\frac{V}{2}\sum_{i\neq j}{\frac{n_i n_j}{|i-j|^3}},
\end{equation}
with $\alpha=1/3$. Under the periodic boundary condition, we can
transform the Hamiltonian from the real space to the momentum
space through a Fourier transformation. For $\alpha=1/3$, there
are 3 sites in each unit cell and the single-particle energy
spectrum are composed of 3 bands. For convenience, we use
$\gamma_{1i}$, $\gamma_{2i}$, and $\gamma_{3i}$ to represent the
annihilation operator at different sites in a unit cell,
respectively. To represent it in momentum space, we make the
Fourier transformation
\begin{equation}
\label{eqn2} \gamma_{\beta
m}=\frac{1}{\sqrt{N_{cell}}}\sum_k{\gamma_{\beta k}e^{-ikx_m}}
\end{equation}
 where $\beta=1,2,3$, $k=2\pi l/N_{cell}$, $l=0,1\ldots N_{cell}-1$ and m is the
 index of the unit cell. So the Hamiltonian $H=H_0 + H_{int}$ can be represented in momentum space,
and the noninteracting part $H_0$ is given by
\begin{eqnarray}
\label{eqn3}
H_0&=&\sum_k{(\gamma_{1k}^{\dag}\gamma_{2k}+\gamma_{2k}^{\dag}\gamma_{3k}+\gamma_{3k}^{\dag}\gamma_{1k}e^{-ik}
+\mathrm{H.c.})}+\sum_{k,\beta}{\lambda\cos{(2\pi
\beta/3+\delta)}\gamma_{\beta k}^{\dag}\gamma_{\beta k}}.
\end{eqnarray}
The interacting part $H_{int}$ can be represented as $H_{int}=
\sum_{\beta\leq\beta'} H_{int}^{\beta\beta'}$ with $\beta$ and
$\beta'$ taking $1$, $2$ and $3$. For the case of $N_{cell}=even$,
\begin{eqnarray}
\label{eqn4}
H_{int}^{\beta\beta}&=&\sum_{k_1,k_2,k_3,k_4}\{\frac{V}{N_{cell}}\sum_{\eta=1}^{N_{cell}/2}{\frac{\cos[(k_3-k_4)\eta]}{(3\eta)^3}}
-\frac{V}{2N_{cell}}\frac{\cos{\frac{(k_3-k_4)N_{cell}}{2}}}{(\frac{3N_{cell}}{2})^3}\}
\gamma_{\beta k_1}^{\dag}\gamma_{\beta k_2}\gamma_{\beta k_3}^{\dag}\gamma_{\beta k_4}\delta_{(k_1-k_2+k_3-k_4)\mathrm{mod}(2\pi)}, \nonumber \\
H_{int}^{12}&=&\frac{V}{N_{cell}}\sum_{k_1,k_2,k_3,k_4}\sum_{\eta=0}^{N_{cell}/2-1}[\frac{e^{i(k_3-k_4)\eta}}{(3\eta+1)^3}
+\frac{e^{-i(k_3-k_4)(\eta+1)}}{(3\eta+2)^3}]\gamma_{1k_1}^{\dag}\gamma_{1k_2}\gamma_{2k_3}^{\dag}\gamma_{2k_4}\delta_{(k_1-k_2+k_3-k_4)\mathrm{mod}(2\pi)}, \nonumber \\
H_{int}^{13}&=&\frac{V}{N_{cell}}\sum_{k_1,k_2,k_3,k_4}\sum_{\eta=0}^{N_{cell}/2-1}[\frac{e^{i(k_3-k_4)\eta}}{(3\eta+2)^3}
+\frac{e^{-i(k_3-k_4)(\eta+1)}}{(3\eta+1)^3}]\gamma_{1k_1}^{\dag}\gamma_{1k_2}\gamma_{3k_3}^{\dag}\gamma_{3k_4}\delta_{(k_1-k_2+k_3-k_4)\mathrm{mod}(2\pi)}.
\nonumber
\end{eqnarray}
For the case of $N_{cell}=odd$,
\begin{eqnarray}
\label{eqn5}
H_{int}^{\beta\beta}&=&\frac{V}{N_{cell}}\sum_{k_1,k_2,k_3,k_4}\sum_{\eta=1}^{(N_{cell}-1)/2}{\frac{\cos[(k_3-k_4)\eta]}{(3\eta)^3}}
\gamma_{\beta k_1}^{\dag}\gamma_{\beta k_2}\gamma_{\beta k_3}^{\dag}\gamma_{\beta k_4}\delta_{(k_1-k_2+k_3-k_4)\mathrm{mod}(2\pi)}, \nonumber \\
H_{int}^{12}&=&\frac{V}{N_{cell}}\{\sum_{k_1,k_2,k_3,k_4}\sum_{\eta=1}^{(N_{cell}-1)/2}[\frac{e^{i(k_3-k_4)\eta}}{(3\eta+1)^3}
+\frac{e^{-i(k_3-k_4)\eta}}{(3\eta-1)^3}]+1\}\gamma_{1k_1}^{\dag}\gamma_{1k_2}\gamma_{2k_3}^{\dag}\gamma_{2k_4}\delta_{(k_1-k_2+k_3-k_4)\mathrm{mod}(2\pi)}, \nonumber \\
H_{int}^{13}&=&\frac{V}{N_{cell}}\{\sum_{k_1,k_2,k_3,k_4}\sum_{\eta=1}^{(N_{cell}-1)/2}[\frac{e^{i(k_3-k_4)(\eta-1)}}{(3\eta-1)^3}
+\frac{e^{-i(k_3-k_4)(\eta+1)}}{(3\eta+1)^3}]+e^{-i(k_3-k_4)}\}\gamma_{1k_1}^{\dag}\gamma_{1k_2}\gamma_{3k_3}^{\dag}\gamma_{3k_4}
\delta_{(k_1-k_2+k_3-k_4)\mathrm{mod}(2\pi)}. \nonumber
\end{eqnarray}
Here $H_{int}^{\beta_1\beta_2}$ is for long-range interactions
between sites $\beta_1$ and $\beta_2$ in the Hamiltonian, and
$\beta_1$, $\beta_2$ are for the indexes of the sites in a unit
cell. We note that $H_{int}^{23}$ and $H_{int}^{12}$ have the same
form for both cases. Under the periodic boundary condition, total
quasi-momentum of the system $\hat{K}
=\sum_{l=0}^{N_{cell}-1}\sum_{\beta=1,2,3}(2\pi
l/N_{cell})\gamma_{\beta l}^{\dagger}\gamma_{\beta l}
\mathrm{mod}(2\pi)$ is conservative. It can decompose the total
Hilbert space into subspaces according to the irreducible
representations. The Hamiltonian cannot couple two states
belonging to two different total momentum $\hat{K}$ and thus it is
block diagonalized. We can diagonalize the Hamiltonian in the each
subspace of total momentum $\hat{K}$. Under the twisted boundary
condition, each momentum $k$ in Brillouin zone makes a shift
$k'=k+\theta/N_{cell}$. While the total momentum $\hat{K}$ is
still a good quantum number, the Hamiltonian can be diagonalized
in each subspace of total momentum $\hat{K}$. By changing
$\theta$, the spectrum flux can be obtained.

For the fractionally filling flat band system considered in
\cite{Regnault,Sheng}, the Hamiltonian can be diagonalized in the
filled band by approximately neglecting the other bands, which is
similar to the projecting to the lowest Landau level in
traditional FQH effect \cite{Regnault}. The dimension of the
Hilbert space of the Hamiltonian is thus greatly reduced. In our
case, we take a standard hopping form with a cosine dispersion. It
is hard to ensure the availability of the projection method, due
to the extended band and the large $V$. Taking $N=10$ as an
example, in the current parameters of $\alpha=1/3$ and $\nu=1/3$,
we need choose $L=90$. The projecting method only considers the
filled bands (here is the lowest band), and the dimension of the
Hamiltonian is reduced to about $1.0 \times 10^6$ for each total
momentum sector. While in our case, we need consider all the
bands, the dimension (about $2.0 \times 10^{11}$ for each total
momentum sector) is too huge to calculate by using the full
diagonalization method.

\begin{figure}[tbp]
\includegraphics[height=9cm,width=9cm] {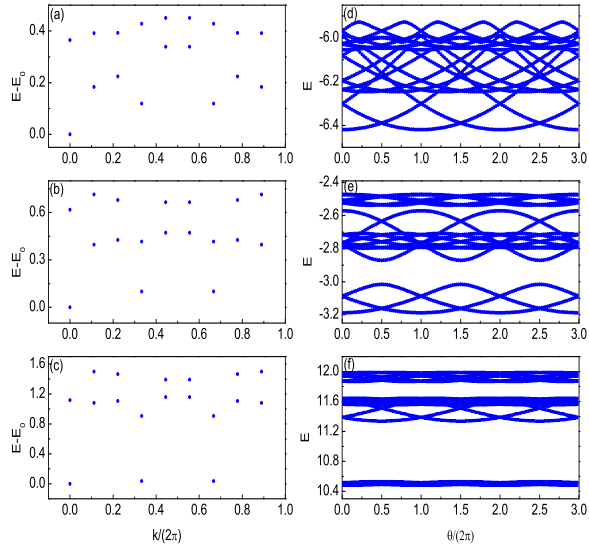}
\caption{(Color online) (a)-(c) Low-energy spectra in the momentum
space for the system with the Coulomb interaction, $t=1,
\lambda=1.5,\alpha=1/3, \delta=5\pi/4$ and different $V$ under the
periodic boundary condition. (d)-(f) Spectrum flow as the function
of $\theta/(2\pi)$ for the system with Coulomb interaction, $t=1,
\lambda=1.5,\alpha=1/3, \delta=5\pi/4$ and different $V$ under the
twist boundary condition. Here, $N=3$ and from top to bottom,
$V=1,10,50$. }\label{sfig1}
\end{figure}

\subsection{II. Case with the Coulomb interaction} In the main
text, we study the case with the long-range dipole-dipole
interaction decaying as $1/r^3$, which corresponds to the dipolar
Fermi gas. We would like to indicate that the main conclusions in
the main text do not depend on the specifical form of
dipole-dipole interaction. For systems with the Coulomb
interaction decaying as $1/r$, we can obtain similar conclusions.
To give an example, we study the system with Coulomb interaction
described by the Hamiltonian:
\begin{equation}
H=-t\sum_{i}{(c_{i}^{\dag}c_{i+1}+\mathrm{H.c.})}+
\sum_{i}{\lambda \cos{(2\pi\alpha i+\delta)}n_i}+
\frac{V}{2}\sum_{i\neq j}{\frac{n_i n_j}{|i-j|}} \label{Coulomb}
\end{equation}
with $\alpha=1/3$. The Hamiltonian can be diagonalized in the
momentum space by following procedures described in the above
section. In Fig.\ref{sfig1}(a)-(c), we show the low-energy
spectrum for the above system with $t=1$, $\lambda=1.5$,
$\alpha=1/3$, $\delta=5\pi/4$, $N=3$ and different $V$ under the
periodic boundary condition. Similar to the dipolar system, when
$V$ is small, for example, $V=1$, the low-energy parts assemble
together and it is hard to distinguish the lowest excited states
apart from the higher excited states by an obvious gap. When
$V=10$, the lowest three states are separated from the higher
excited states with an obvious gap. As $V$ increases further, the
gap becomes more obvious and the lowest three states become nearly
degenerate at $V=50$.

Similar to the dipolar system considered in the main text, we also
calculate the spectrum flow of the system with Coulomb interaction
under twist boundary conditions. In Fig.\ref{sfig1}(d)-(f), we
show the spectrum flow as the function of $\theta$ for the system
described by Eq.(\ref{Coulomb}) with $t=1$, $\lambda=1.5$,
$\alpha=1/3$, $\delta=5\pi/4$, $N=3$ and different $V$ under twist
boundary conditions. As shown in Fig.\ref{sfig1}(d), when $V=1$,
the lower energy levels overlap together with the change of
$\theta$. When $V=10$, the lowest three energy spectra are
separated from the higher excited states with a small gap, while
they flow into each other with the change of $\theta$. For $V=50$,
the lowest three states are nearly degenerate, and the nearly
degenerate ground states are well separated from the higher
excited states by an obvious gap.  In contrast to the dipolar
Fermi system, a smaller $V$ is needed for the emergence of
degenerate ground states due to the Coulomb interaction is much
stronger than the dipole-dipole interaction. Following the
procedure in the main text, we can also verify that the total
Chern number of the three degenerate ground states is $1$ as the
dipolar system. Such an example indicates that the fractional
topological phase can also appear in one-dimensional superlattice
systems with the Coulomb long-range interactions.
\begin{figure}[tbp]
\includegraphics[height=9cm,width=9cm] {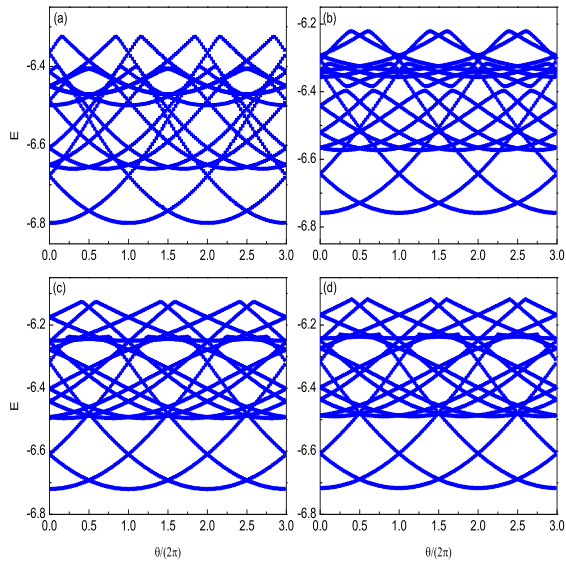}
\caption{(Color online) Spectrum flux as the function of $\theta$
with $t=1, \lambda=1.5,\alpha=1/3, \delta=5\pi/4,N=3$ and
different $R$ under the twist boundary condition. (a)-(c) $V=500$
and $R=1,2,3$, respectively. (d) $V=2000$ and $R=3$.}\label{sfig2}
\end{figure}

\subsection{III. Cases with short-range interactions} In
Ref.\cite{Sheng}, fractional topological states are found in a
two-dimensional lattice system with topological nontrivial flat
band, in which the short-range interaction is used. While in our
system the energy band is split into three bands with finite band
widths, the long-range interaction plays an important role in the
formation of fractional topological states. To see it clearly,  we
consider the Fermi system trapped in the one-dimensional
bichromatic optical lattices with a short-range interaction
\begin{equation}
\label{eqn6} H=-t\sum_{i}{(c_{i}^{\dag}c_{i+1}+\mathrm{H.c.})}+
\sum_{i}{\lambda \cos{(2\pi\alpha i+\delta)}n_i}+
\frac{V}{2}\sum^{\prime}_{|i-j|\leq R}{\frac{n_i n_j}{|i-j|^3}},
\end{equation}
where the last term of Eq.\ref{eqn6} describes the short-range
interaction with $R$ controlling the range of the interaction and
the prime denoting $i\neq j$ due to the Pauli principle. The
Hamiltonian can be also diagonalized under the twist boundary
condition. In order to compare with the dipolar system considered
in the main text, we calculate the spectrum flux of the
Hamiltonian (\ref{eqn6}) with $R=1$, $2$ and $3$. In
Fig.\ref{sfig2}a-Fig.\ref{sfig2}c, we show the spectrum flux for
the system with $V=500$, $t=1$, $\alpha=1/3$, $\delta=5\pi/4$,
$N=3$ and $R=1,2,3$, respectively. In contrast to Fig.3f in the
main text, the lowest three energy levels are not separated with
the higher excited states and no a degenerate ground-state
manifold is formed for $V=500$. Moreover, the lowest three levels
and the exited ones are not well separated even for a much larger
$V=2000$ with $R=3$.
%In large $V$ limits, for the case of $R=3$, there is only one particle in successive 4 sites,
%and the system is equivalent to the $N$ spinless fermions moving at $L-3N$ lattice sites.
As shown in Fig.\ref{sfig2}d, the spectrum flux is almost the same
as the one displayed in Fig.\ref{sfig2}c. Our results show that no
3-fold degenerate ground states can be found and no fractional
topological states emerge for the superlattice system with
short-range interactions even in the strongly interacting limit.
\begin{figure}[tbp]
\includegraphics[height=9cm,width=12cm] {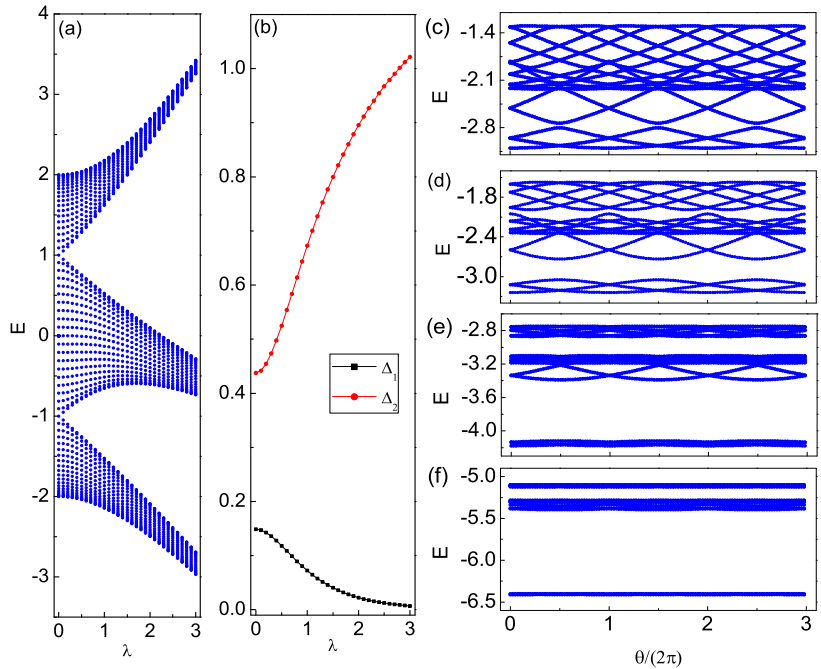}
\caption{(Color online) (a) Single-particle spectrum as a function
of $\lambda$  for system with $t=1$, $\alpha=1/3$,
$\delta=5\pi/4$, and $L=120$ under the periodic boundary
condition. (b) $\Delta_1$ and $\Delta_2$ as the function of
$\lambda$ for the system with $t=1$, $\alpha=1/3$,
$\delta=5\pi/4$, $V=500$, $N=3$ and $L=27$ under the periodic
boundary condition. (c)-(f) Spectrum flux as the function of
$\theta$ for the system with $t=1$, $\alpha=1/3$, $\delta=5\pi/4$,
$V=500$, $N=3$, $L=27$ and different $\lambda$ under the twist
boundary condition. Here, $\lambda=0.1$, $0.5$, $1.5$, $3$ for
(c)-(f), respectively.} \label{sfig3}
\end{figure}

\subsection{IV. Effect of periodical potential} It is worth to
point that the periodically modulated potential provides an
additional dimension to ensure the topologically nontrivial
property in our system. Such a one-dimensional system can be
mapped to the lattice version of the two-dimensional integer Hall
effect (the Hofstadter problem) \cite{Hofstadter} by substituting
$t\to t_x$, $\lambda \to -2t_y$, $\alpha=\Phi$ and $\delta \to
-k_y$ where $\Phi$ is the magnetic flux per unit cell \cite{Lang}.
It is well known that the later model has nontrivial topological
properties. In the main text, we choose the amplitude of the
periodic potential as $\lambda=1.5$. To understand how the
amplitude of $\lambda$ affects our calculation,  we present the
single-particle spectra as a function of $\lambda$ for system with
$t=1$, $\alpha=1/3$, $\delta=5\pi/4$ and $L=120$ under the
periodic boundary condition in Fig.\ref{sfig3}(a). It is shown
that the spectra are split into three bands and the gap between
the first and second band becomes narrow when one decreases
$\lambda$. To get degenerate ground states which are well
separated from the higher excited states, one need choose a
suitable $\lambda$ to generate an obvious gap between the first
and second band. In Fig.\ref{sfig3}(b), we show the width of
lowest three states defined as $\Delta_1=E_2-E_0$ and the energy
gap between the fourth state and third state defined as
$\Delta_2=E_3-E_2$ versus the strength of modulation potential
$\lambda$
for the system with $t=1$, $\alpha=1/3$, $\delta=5\pi/4$, $V=500$ and $N=3$, where $E_0$, $E_1$, $E_2$ and $E_3$ denote energies of the lowest four eigenstates. %In Fig.\ref{sfig3}(b), the square is for
%the bandwidth $\Delta_1$ and the circle is for the energy gap $\Delta_2$.
From Fig.\ref{sfig3}(b), it can be seen that $\Delta_1$ decreases
whereas $\Delta_2$ increases with the increase of $\lambda$. In
the limit of $\lambda \to 0$, we can find that $\Delta_1$ and
$\Delta_2$ have the same scale, which is not favorable for the
formation of degenerate ground states. Fig.\ref{sfig3}(c)-(f) show
the spectrum flux for the system with $t=1$, $\alpha=1/3$,
$\delta=5\pi/4$, $V=500$, $N=3$ and different $\lambda$. For
$\lambda=0.1$, the lowest three states are not well separated from
the higher states. With the increase of $\lambda$, the gap becomes
more obvious. For $\lambda=1.5$ chosen in the main text for the
calculation, the lowest three states are nearly degenerate with an
obvious gap separated them from the higher states.


\begin{references}

\bibitem{Cooper} N. R. Cooper, Adv. Phys. {\bf 57}, 539 (2008).

\bibitem{Fetter} A. L. Fetter, Rev. Mod. Phys. {\bf 81}, 647 (2009).

\bibitem{Spielman} Y.-J. Lin, R. L. Compton, K. Jimnez-Garcia, J.V. Porto, and I. B.
Spielman, Nature (London) {\bf 462}, 628 (2009).

%\bibitem{Wilkin} N. K. Wilkin and J. M. F. Gunn, Phys. Rev. Lett. 84, 6 (2000); B.
%Paredes, P. Fedichev, J. I. Cirac, and P. Zoller, Phys. Rev. Lett.
%87, 010402 (2001).

\bibitem{Osterloh07} K. Osterloh, N. Barber\'{a}n, and M. Lewenstein, Phys. Rev. Lett. {\bf 99},
160403 (2007).

%\bibitem{Baranov} M. A. Baranov, H. Fehrmann, and M. Lewenstein, Phys. Rev. Lett.
%100, 200402 (2008).
%Wigner Crystallization in Rapidly Rotating 2D Dipolar Fermi Gases

\bibitem{Qiu}  R.-Z. Qiu, F. D. M. Haldane, X. Wan, K. Yang, and S. Yi
, Phys. Rev. B {\bf 85}, 115308 (2012); R.-Z. Qiu, S.-P. Kou,
Z.-X. Hu, Xin Wan, and S. Yi, Phys. Rev. A {\bf 83}, 063633 (2011).



\bibitem{Osterloh05} K. Osterloh, M. Baig, L. Santos, P. Zoller, and M. Lewenstein,
Phys. Rev. Lett. {\bf 95}, 010403 (2005).
%Cold Atoms in Non-Abelian Gauge Potentials: From the Hofstadter
%"Moth" to Lattice Gauge Theory

\bibitem{Shao}
L. B. Shao, S.-L. Zhu, L. Sheng, D. Y. Xing, and Z. D. Wang, Phys.
Rev. Lett. {\bf 101}, 246810 (2008).
%Realizing and Detecting the Quantum
%Hall Effect without Landau Levels by Using Ultracold Atoms

\bibitem{Kane} M. Z. Hasan and C. L. Kane, Rev. Mod. Phys. \textbf{82}, 3045
(2010); X.-L. Qi and S.-C. Zhang,  Rev. Mod. Phys. \textbf{83},
1057 (2011).

\bibitem{Lang} L.-J. Lang, X. M. Cai, and S. Chen, Phys. Rev. Lett. {\bf 108},
220401 (2012).

\bibitem{Kraus} Y. E. Kraus, Y. Lahini, Z. Ringel, M. Verbin, and O.
Zilberberg, Phys. Rev. Lett. {\bf 109}, 106402 (2012).

\bibitem{Zhu} F. Mei, S. L. Zhu, Z. M. Zhang, C. H. Oh, and N. Goldman Phys. Rev. A {\bf 85}, 013638 (2012).

\bibitem{TKNN} D. J. Thouless, M. Kohmoto, M. P. Nightingale, and M. den Nijs,
Phys. Rev. Lett. \textbf{49}, 405 (1982).

\bibitem{Hofstadter} D. R. Hofstadter, Phys. Rev. B \textbf{14}, 2239 (1976).

\bibitem{exp1} L. Fallani, J. E. Lye, V. Guarrera, C. Fort, and M.
Inguscio, Phys. Rev. Lett. \textbf{98}, 130404 (2007).
\bibitem{exp2} G. Roati \emph{et al}., Nature (London) \textbf{453},
895 (2008); B. Deissler \emph{et al}., Nat. Phys. \textbf{6}, 354
(2010).

\bibitem{dipolar} K.-K. Ni, S. Ospelkaus, M. H. G. de Miranda,
A. Pe'er, B. Neyenhuis, J. J. Zirbel, S. Kotochigova, P. S.
Julienne, D. S. Jin, and J. Ye, Science \textbf{322}, 231 (2008);
S. Ospelkaus, A. Pe'er, K.-K. Ni, J. J. Zirbel, B. Neyenhuis, S.
Kotochigova, P. S. Julienne, J. Ye, and D. S. Jin, Nat. Phys.
\textbf{4}, 622 (2008).

\bibitem{suppl} See Supplemental Material for details.

\bibitem{Regnault} N. Regnault and B. A. Bernevig, Phys. Rev. X {\bf 1}, 021014
(2011).

\bibitem{Bergholtz} E. J. Bergholtz and A. Karlhede, Phys. Rev. B {\bf 77},
155308 (2008).


\bibitem{Laughlin} R. B. Laughlin, Phys. Rev. Lett. {\bf 50}, 1395 (1983).

\bibitem{Halperin} B. I. Halperin, Phys. Rev. Lett. {\bf 52}, 1583, (1984);
D. Arovas, J. R. Schrieffer, and F. Wilczek, Phys. Rev. Lett.
{\bf 53}, 722 (1984).

\bibitem{Niu} Q. Niu, D. J. Thouless, and Y. S. Wu, Phys. Rev. B {\bf 31},
3372 (1985).

\bibitem{Sheng} D. N. Sheng, Z. C. Gu, K. Sun and L. Sheng, Nat. Commun. {\bf 2}, 389
(2011).

\bibitem{Seidel}  A. Seidel, H. Fu, D.-H. Lee, J. M. Leinaas, and J. Moore,
Phys. Rev. Lett. {\bf 95}, 266405 (2005).

%\bibitem{Bergholtz} E. J. Bergholtz and A. Karlhede, J. Stat. Mech. (2006) L04001.

%\bibitem{DHLee} D. H. Lee and J.M. Leinaas, Phys. Rev. Lett. {\bf 92}, 096401 (2004).

\bibitem{Bergholtz-PRL} E. J. Bergholtz and A. Karlhede, Phys. Rev. Lett. {\bf 94}, 026802
(2005).

\bibitem{Bergholtz2} E. J. Bergholtz and A. Karlhede, J. Stat. Mech. (2006) L04001; A. Seidel and D.-H. Lee,
Phys. Rev. Lett. {\bf 97}, 056804 (2006).


%\bibitem{Wang} Yi-Fei Wang, Zheng-Cheng Gu, Chang-De Gong, and D. N. Sheng
%Phys. Rev. Lett. {\bf 107}, 146803 (2011).
%\bibitem{D Wang} D. Wang, B. Neyenhuis, M. H. G. de Miranda, K.-K. Ni, S. Ospelkaus,
%D. S. Jin, and J. Ye, Phys. Rev A \textbf{81}, 061404 (2010).

\end{references}
\end{document}